\documentclass{optica-article}

\journal{opticajournal}

\articletype{Research Article}

\begin{document}

\title{Characterization and evasion of backscattered light in the squeezed-light enhanced gravitational wave interferometer GEO\,600}

\author{Fabio Bergamin,\authormark{1,*} James Lough,\authormark{1} Emil Schreiber,\authormark{1} Hartmut Grote, \authormark{2} Moritz Mehmet, \authormark{1} Henning Vahlbruch, \authormark{1} Christoph Affeldt, \authormark{1} Tomislav Andric, \authormark{1} Aparna Bisht, \authormark{1} Marc Bringmann, \authormark{1} Volker Kringel, \authormark{1} Harald L\"uck, \authormark{1} Nikhil Mukund, \authormark{1} Severin Nadji, \authormark{1} Borja Sorazu, \authormark{3} Kenneth Strain, \authormark{3} Michael Weinert, \authormark{1} and Karsten Danzmann \authormark{1}}

\address{\authormark{1}Max-Planck-Institut f\"ur Gravitationsphysik (Albert-Einstein-Institut) and Leibniz Universit\"at Hannover, Callinstr.\,38, 30167 Hannover, Germany\\
\authormark{2}School of Physics and Astronomy, Cardiff University, The Parade, CF24 3AA, United Kingdom\\
\authormark{3}SUPA, University of Glasgow, Glasgow G12 8QQ, United Kingdom}

\email{\authormark{*}fabio.bergamin@aei.mpg.de}

\begin{abstract*} 
Squeezed light is injected into the dark port of gravitational wave interferometers, in order to reduce the quantum noise. 
A fraction of the interferometer output light can reach the OPO due to sub-optimal isolation of the squeezing injection path.
This backscattered light interacts with squeezed light generation process, introducing additional measurement noise. We present a theoretical description of the noise coupling mechanism.\\
\noindent We propose a control scheme to achieve a de-amplification of the backscattered light inside the OPO with a consequent reduction of the noise caused by it. The scheme was implemented at the GEO\,600 detector and has proven to be crucial in maintaining a good level of quantum noise reduction of the interferometer for high parametric gain of the OPO. In particular, the mitigation of the backscattered light noise helped in reaching $6\,\mathrm{dB}$ of quantum noise reduction [Phys. Rev. Lett. {\bfseries 126}, 041102 (2021)].\\
\noindent The impact of backscattered-light-induced noise on the squeezing performance is phenomenologically equivalent to increased phase noise of the squeezing angle control. The results discussed in this paper provide a way for a more accurate estimation of the residual phase noise of the squeezed light field.
\end{abstract*}

\section{Introduction}
\begin{figure}[ht]
    \centering
    \includegraphics[width=1\textwidth]{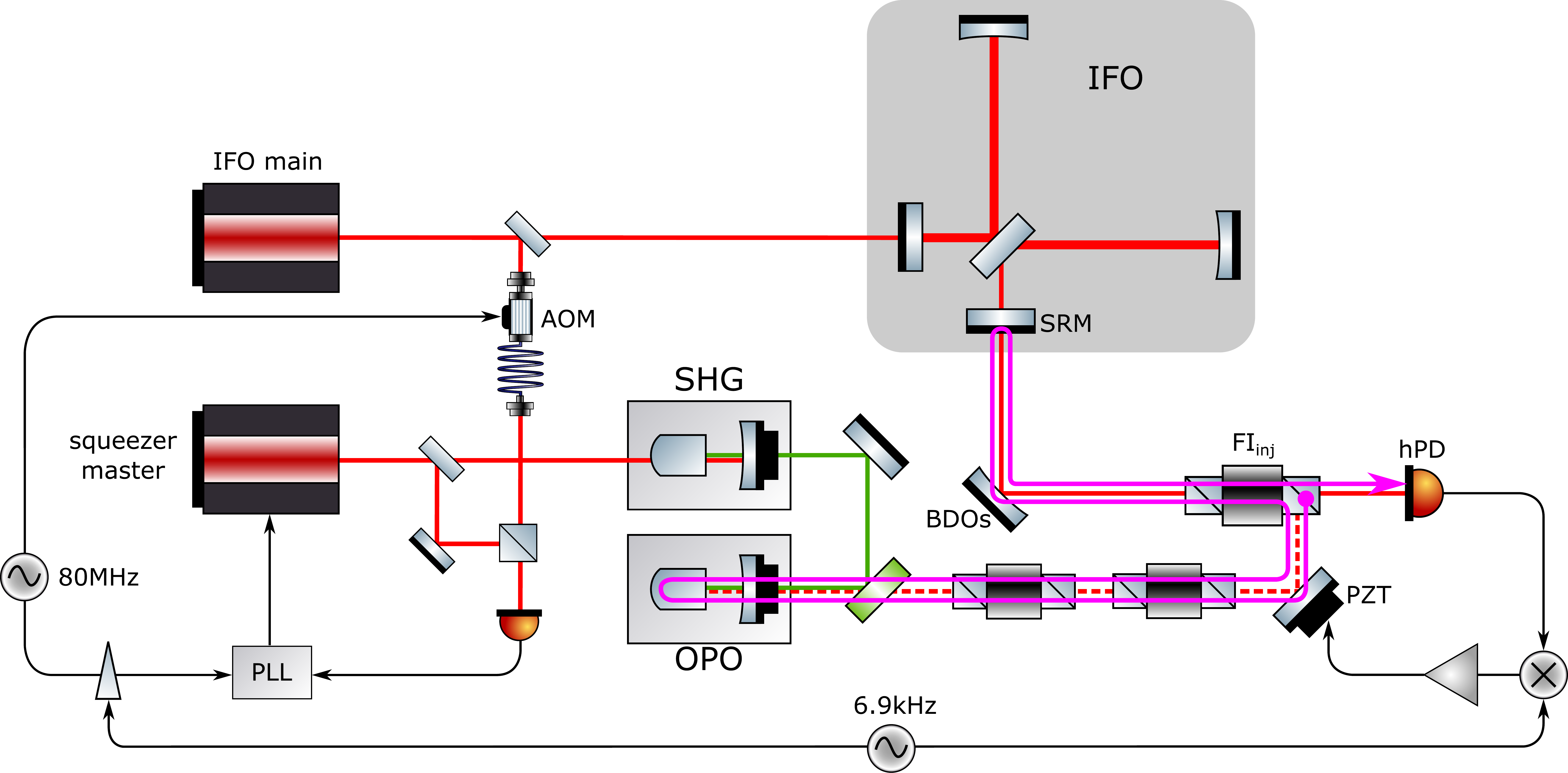}
    \caption
    {Overview of the squeezing injection subsystem. The squeezer master laser is phase-locked to the main laser of the IFO through a PLL at $80\,\textrm{MHz}$. The pick-off of the main laser is frequency shifted by $80\,\textrm{MHz}$ through an acousto-optic modulator (AOM) at the input of the fiber, thus allowing a phase lock at the same frequency between the two lasers. The squeezed light and the pump field are indicated by the dashed red line and by the green line, respectively. The purple arrow indicates the path traveled by the backscattered light, starting from the imperfect PBS of the injection Faraday rotator. The schematic shows also the main features of the backscatter phase control system described in section \ref{sec:BSP_lock}: the squeezed light phase is dithered at 6868Hz via the PLL's LO. The actuation, obtained through the FM input of the LO, is represented as a phase shifter in the figure. The resulting line in the hPD signal is then demodulated and fed to a piezo-mounted mirror (PZT) after filtering, counteracting in this way the path-length fluctuation between the signal-recycling mirror (SRM) and the OPO. This ensures a phase lock condition between the squeezing phase and the backscattered-light phase.}
    \label{fig:schematic}
\end{figure}

In the framework of gravitational wave detectors \cite{LIGOScientific:2014pky, VIRGO:2014yos,Luck:2010rt,Aso:2013eba}, quantum-noise-limited interferometric measurements are very sensitive to the influence of scattered light. Stray light that reaches the photodetectors by another but the intended path will generally have randomly fluctuating phase and amplitude and will add a noisy contribution to the measured signal.\\
\indent Typical scattering sources are, for example, small imperfections of optical surfaces, like point absorbers \cite{PhysRevLett.127.241102}, and residual reflections from AR coatings.
Scattered light coming from the environment, such as reflections from the vacuum envelope \cite{Vinet1997} or from baffles \cite{pelosi2017baffle}, has in general less impact. The propagation direction of this light does not coincide with the axis of the Gaussian beam carrying the signal, and just a small fraction of this stray light is scattered off of the optic surfaces back into the interferometer (IFO) mode \cite{Ottaway2012}. 
In the special case of a squeezing-enhanced IFO  \cite{PhysRevD.23.1693}, the squeezed light source itself can become an important scatterer.\\
\indent The squeezed vacuum field is injected at the output port of the IFO with the help of a Faraday rotator ($\mathrm{FI_{inj}}$ in fig.~\ref{fig:schematic}), but polarization imperfections can lead to some light from the IFO output being sent backwards along the squeezing-injection path.
Stray light hitting the parametric amplifier is then re-injected into the IFO, creating unwanted phase fluctuations at the readout photodiode (hPD). This paper is focused on this kind of stray light, which will be called as backscattered light.
To describe this process, special care must be taken to consider the nonlinear interaction of the stray-light field with the parametric amplifier.
Of particular interest is the back-propagating light in the same polarization and spatial mode as the squeezed field because when scattered back toward the IFO, light in this mode will reach the main photodetector with high efficiency.\\
The squeezed light sources currently used in gravitational wave interferometers are called optical parametric oscillators (OPO). These are optical cavities with an embedded second-order nonlinear medium. The geometry of the OPO may vary for different detectors, as in the case of Virgo \cite{acernese2019increasing} and LIGO \cite{tse2019quantum}.
At GEO\,600 \cite{Luck:2010rt}, the OPO is a hemilithic cavity containing a plano-convex PPKTP crystal of about 10 mm length \cite{Vahlbruch_2010}. 
For the back-propagating stray-light field, the OPO is a highly over-coupled resonant cavity, which means that most of the field enters the cavity where it experiences the parametric gain before being reflected back toward the IFO.
This can also be understood as seeding the OPO with the bright stray-light field instead of a vacuum field, thus creating a displaced squeezed state.\\
\indent This paper is focused on the coupling of scattered light noise after the back-reflection off the OPO. Section \ref{sec:description_of_bsc} provides an analytical model of the parametric interaction. Experimental results confirming this model are presented in section \ref{sec:experimental_results}. A new technique, developed at GEO\,600, to actively control the backscattered light phase and reduce its contribution to the final noise is introduced in section \ref{sec:BSP_lock}. In section \ref{sec:influence_phase_noise_est}, finally, the impact of the new technique on the estimation of squeezed-light phase noise is discussed.

\section{Description of backscatter coupling}\label{sec:description_of_bsc}
The approach adopted in the following can be considered semi-classical since the extra-cavity fields that solve the equations of motion for the nonlinear interaction are described by complex numbers: the steady-state component of the linearization \cite[\S3.1.5]{white1997classical}.
Reference \cite{mckenzie2008squeezing} describes a detailed model of the noise coupling using quantum operators but considering the assumption of constant phase between pump and seed field.\\
Furthermore, we consider a slowly-varying approximation: the equations of motion are solved for a steady state, but the seed is free to change its phase in the phasor diagram. However, the rotation rate is much slower than the cavity round-trip time, so at each measurement time, the phase relation between the seed and pump field can be considered fixed, and the cavity equations can be solved for the steady-state case.\\
Solving the steady-state equations of motion for a seeded OPO \cite[\S2.4.7]{Goda2007} leads to the following expression for the reflected bright field in a reference frame rotating at the optical frequency:

\begin{equation}
    \mathrm{A_{bsc,ref}}=\mathrm{A_{bsc,in}}e^{i\phi}\left( 2\eta_{esc}\frac{1-xe^{2i(\theta-\phi)}}{1-x^2}-1 \right)\;,
\end{equation}
where $\eta_{esc}$ is the escape efficiency, $x=\sqrt{\mathrm{P_{pump}/P_{thres}}}<1$ is the normalized nonlinear interaction strength that can be expressed in terms of pump power $\mathrm{P_{pump}}$ and the pump power required to reach the optical parametric oscillation threshold $\mathrm{P_{thres}}$. $\mathrm{A_{bsc,in}}$ and $\phi$ are the amplitude and the randomly fluctuating phase of the backscattered light, respectively, and $\theta$ is the squeezing angle. The above expression embeds the parametric amplification process: the reflected field is amplified or deamplified according to the phase relation with the pump field (note that in this reference system, the pump field phase is $2\theta$).\\
The bright squeezed field is injected into the IFO and eventually reaches hPD, interfering with the output light from the IFO. Let the IFO output light amplitude be described by the real value $\mathrm{A_{IFO}}$ so that the other phases are referred to its reference frame. Assuming the backscattered light field to be small with respect to this field, the noise signal at the hPD is:

\begin{align}
    \mathrm{I_{bsc}} &\propto \mathrm{A_{IFO}}\Re[\mathrm{A_{bsc,ref}}]\nonumber\\
    &= \mathrm{A_{IFO}A_{bsc,in}}\left[
    \left(-1+\frac{2\eta_{esc}}{1-x^2}\right)\cos(\phi)-\left(\frac{2\eta_{esc} x}{1-x^2}\right)\cos(2\theta-\phi)\right]\;.
    \label{eq:backscatter_signal_raw}
\end{align}
The phases can be split into two parts

\begin{align}
    \begin{split}
        \phi &= \Phi + \delta\phi
        \;,\\[1em]
        \theta &= \Theta + \delta\theta
        \;,
    \end{split}
\end{align}
where $\Phi(t)$ and $\Theta(t)$ are the slow varying contributions---at frequencies below the detection band---and $\delta\phi(t)$ and $\delta\theta(t)$ are the fast and small contributions. The squeezing angle is actively controlled to match the squeezed quadrature to the readout quadrature: $\Theta = 0$. Thus Eq.~(\ref{eq:backscatter_signal_raw}) can be expressed as:
\begin{equation}
    \mathrm{I_{bsc}}\propto \mathrm{A_{IFO}A_{bsc,in}}\left[\left(-1+\frac{2\eta_{esc}}{1\pm x}\right)\cos\Phi
    -\left(-1+\frac{2\eta_{esc}}{1\pm x}\right)\sin\Phi\delta\phi
    \mp\frac{4\eta_{esc} x}{1-x^2}\sin\Phi\delta\theta\right]\;,
    \label{eq:backscatter_signal}
\end{equation}
where quadratic terms in $\delta\phi$ and $\delta\theta$ were neglected. The above expression takes into account also the case where antisqueezing ($\Theta=\pi/2$) is injected into the IFO, this corresponds to the lower signs.
Alternatively, the above expression can be re-written in terms of the squeezing factor $r$ by substituting
\begin{equation}
    e^{-r}=\frac{1-x}{1+x}\;,
    \label{eq:pure_state}
\end{equation}
which is the squeezing ratio for a pure squeezed state, $r<0$ means that antisqueezing is applied.
To further simplify Eq. (\ref{eq:backscatter_signal}), $\eta_{esc}\simeq1$ can be assumed, being in general the power loss due to the OPO escape efficiency on the order of $1\%$ \cite{Vahlbruch_2010,tse2019quantum,acernese2019increasing}. This yields:
\begin{equation}
    \mathrm{I_{bsc}}\propto\mathrm{A_{IFO}A_{bsc,in}}[ e^{-r} \cos\Phi
- e^{-r}\sin\Phi\delta\phi
-2\sinh(r)\sin\Phi\delta\theta]
 \label{eq:backscatter_signal_r}\;.
\end{equation}
Eq.~(\ref{eq:backscatter_signal}) and (\ref{eq:backscatter_signal_r}) show that the noise due to backscattered light is made up of three contributions:
\begin{itemize}
\item The first term describes a nonlinear coupling of the slow phase changes $\Phi(t)$ of the stray-light field.
This can be neglected as long as the absolute magnitude of these low-frequency fluctuations is not large enough to up-convert them into the detection bandwidth. At GEO\,600, the calibrated detection bandwidth starts at $40\mathrm{Hz}$. A more detailed explanation of the up-convertion process is provided in section \ref{sec:up_convertion}.
\item The second term describes a linear coupling of the small-amplitude phase fluctuations $\delta\phi(t)$ of the stray-light field.
\item The third term, finally, shows a linear coupling of residual squeezing-angle fluctuations $\delta\theta(t)$ in the presence of a stray-light field.
\end{itemize}
It is interesting to note that the first two noise coupling mechanisms actually get smaller when applying more squeezing: the parametric amplification process in the OPO deamplifies the backscattered light field (along with the vacuum fluctuations) in the correct quadrature to reduce its influence on the detected signal. The coupling of squeezing-angle fluctuations, however, increases with the pump power.\\
Eq.~(\ref{eq:backscatter_signal_r}) can be expressed as a power spectral density in units of relative intensity noise at the hPD:
\begin{equation}
    \mathrm{RIN}_\mathrm{bsc}(f) \approx \frac{4\eta_{inj} \mathrm{P_\textup{bsc,in}}}{\mathrm{P_\textup{IFO}}}\left[\frac{1}{2}e^{-2r} S_{\delta\phi}(f) + 2\sinh^2(r) S_{\delta\theta}(f)\right]
    \;,
    \label{eq:backscatter_PSD}
\end{equation}
where $S_{\delta\phi}$ and $S_{\delta\theta}$ are the power-spectral densities (PSDs) of the randomly fluctuating angles $\delta\phi$ and $\delta\theta$, $\mathrm{P_\textup{IFO}}$ is the carrier-light power at the IFO output reaching the hPD, $\mathrm{P_\textup{bsc,in}}$ is the amount of backscattered light in the fundamental mode impinging on the OPO. and $\eta_{inj}$ is the injection efficiency of the backscattered light. This term takes into account effective optical losses, mode mismatch and misalignment with respect to the IFO and to the output mode cleaner (OMC), quantum efficiency and electronic dark noise of the final hPD.
Since $\mathrm{P_\textup{bsc,in}}\propto\mathrm{P_{IFO}}$, the above expression does not depend on power at all, this means that amplitude fluctuations of the backscattered light do not add noise.
In the above formula, the nonlinear coupling term was dropped and the slow phase fluctuations were
averaged over all possible values of $\Phi$.
Since the phase noise depends on the length changes of the injection path, which is traveled by both the fields, $S_{\delta\phi}(f)$ and $S_{\delta\theta}(f)$ will be of similar magnitude outside of the control bandwidth of the squeezing-angle control loop (4kHz). For the same reason, the two effects are partially correlated; thus Eq.~(\ref{eq:backscatter_PSD}) might slightly over-estimate the combined noise.\\
\indent The fact that the coupling of squeezing-angle fluctuations increases with increasing parametric gain leads to a situation where there is a maximum amount of squeezing that can be applied before the added backscatter noise outweighs the reduction of quantum noise. It is evident that this effect becomes a limiting factor if higher squeezing levels are desirable.
Earlier theoretical and experimental investigations \cite{Dwyer2013a, Chua2014, Oelker2014b} of backscattered light noise have described the coupling of the stray light’s phase fluctuations, but have generally assumed the squeezing angle to be constant, thus neglecting the coupling mechanism described by the last term of Eq.~(\ref{eq:backscatter_signal_r}).

\section{Experimental results}\label{sec:experimental_results}
The different coupling mechanisms described by Eq.~(\ref{eq:backscatter_signal})
and  their dependence on the parametric gain are investigated in this section.

\subsection{Coupling of slow phase fluctuations}\label{sec:up_convertion}

\begin{figure}[ht]
    \centering
    \includegraphics[width=1\textwidth]{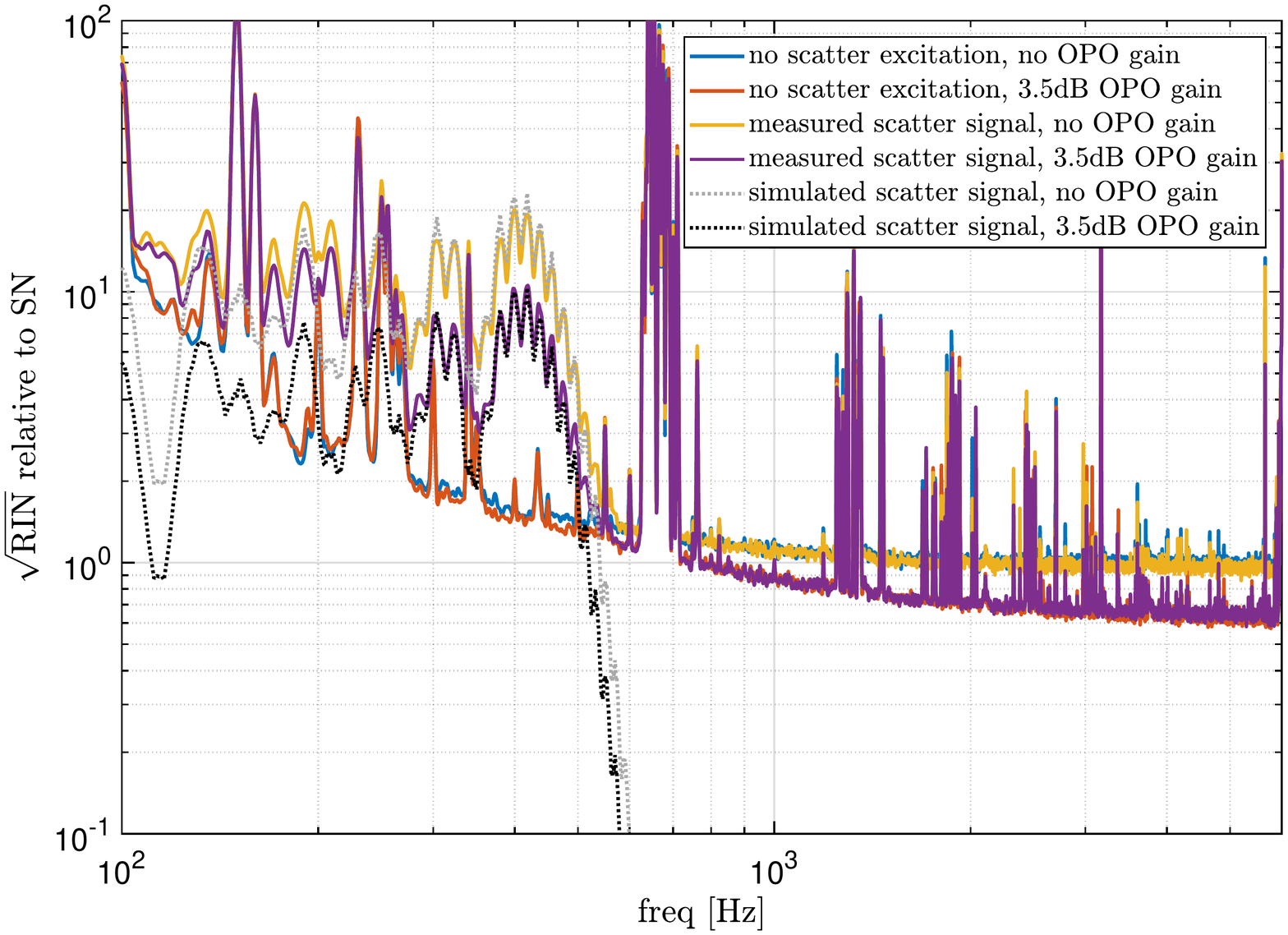}
    \caption{Backscatter noise in the IFO output signal with an intentional large-amplitude excitation of the injection path length (yellow and purple traces). The characteristic `scattering shoulder' is visible and is well modeled by a numerical time-domain simulation (dotted traces). When squeezing is applied, the backscattered noise is de-amplified by the squeezing factor as predicted. The applied modulation had an amplitude of $23.5\,\mathrm{rad}$ at a frequency of $19\,\mathrm{Hz}$.}
    \label{fig:shoulder}
\end{figure}

The effects of backscattering are observed by artificially introducing low-frequency path-length modulations of the squeezing injection path with the help of a mirror mounted on a longitudinal piezoelectric actuator.
For sufficiently large amplitudes, larger than one wavelength, the nonlinear coupling---described by the first term in Eq.~(\ref{eq:backscatter_signal})---can up-convert the slow phase fluctuations into the detection band.
This effect can be understood by considering the Jacobi-Anger expansion of the phase-modulated signal:
\begin{equation}
    e^{iz\cos(\Omega t)}=\sum_{n=-\infty}^\infty i^n J_n(z)e^{in\Omega t},
\end{equation}
where $z$ is the modulation depth, $J_n(z)$ is the n-th Bessel  function. The result is the characteristic `scattering shoulder' in the spectrum of the detector output signal, as shown in 
Fig.~\ref{fig:shoulder}.
The measurement was done both with and without pumping the OPO.
As predicted by Eq.~(\ref{eq:backscatter_signal}), the amount of backscattering depends on the OPO's nonlinear gain.
With the squeezing angle tuned for reducing the shot-noise, the backscatter noise is also attenuated by the same factor.
The plot shows also the simulated scattering shoulder, obtained using
\begin{equation}\label{eq:shoulder}
    \frac{\mathrm{RIN}_\mathrm{shoulder}}{\mathrm{RIN}_\mathrm{sn}}=e^{-2r}\frac{4\eta_{inj}\mathrm{P_{bsc}}}{2\hbar\omega} S_{\cos\Phi}(f),
\end{equation}
which is already scaled in units of shot noise level. $S_{\cos\Phi}$ is the PSD of the large-amplitude excitation.\\
As already pointed out, applying higher squeezing factors only improves backscattering due to stray-light phase fluctuations. Thus, an upper limit on the impact of this form of backscattering can be set by comparing the detector's output signal in the state where stray light reaches the OPO without any pump applied to a situation where the squeezer path is completely blocked with a high-quality beam damp.
For GEO\,600 no difference in the measured noise level was observed, showing that stray-light phase fluctuations are not a limiting source of noise at any frequency within the detection band.
Experiments for the squeezing demonstration at LIGO Hanford had come to similar conclusions but could also show that for the more demanding requirements of future advanced detectors the backscatter immunity was not yet adequate at low frequencies \cite{Chua2014}. This indicates that better optical isolation or more path-length stability will be needed at some point.

\subsection{Coupling of fast phase fluctuations}

Another form of backscatter-induced detection noise is the linear coupling of squeezing angle-fluctuations as described by the third term of Eq.~(\ref{eq:backscatter_signal}).
Experimental evidence for this effect had first been noticed at GEO\,600 when an unexpected degradation of the squeezing level was observed for high nonlinear-gain settings of the OPO.
Initially this was attributed to excess RMS phase noise of the squeezer which would lead to similar observations (see section \ref{sec:influence_phase_noise_est}).
Backscattering was found to be the underlying mechanism when it became clear that the problem had started due to a lowered isolation of one of the Faraday isolators after a rerouting of the path trough the respective optics.
Subsequent experiments could confirm this by intentionally lowering the backscatter isolation and observing the effect on the squeezing level.

\begin{figure}[ht]
    \centering
    \includegraphics[width=1\textwidth]{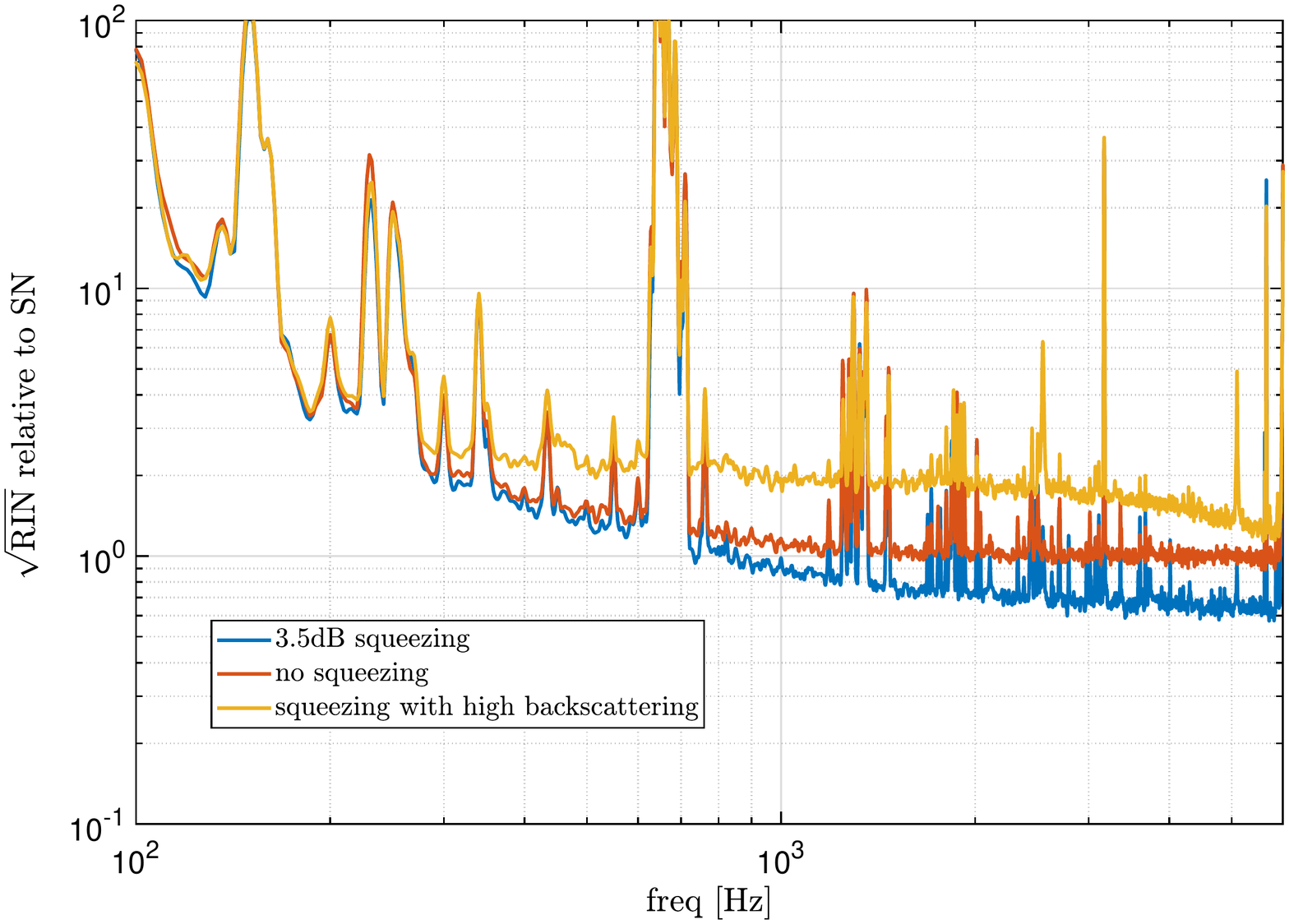}
    \caption{
    RIN of hPD in shot noise level units for good (blue trace) and bad (yellow trace) backscatter isolation. In both the traces, squeezing was applied. The squeezing level measured from the shot noise reduction respect to the reference (red trace) in the nominal situation was $3.5\,\textrm{dB}$. The reduced backscatter isolation was achieved by offsetting one of the PBSs in the in-air squeezing injection path by approximately $0.06^\circ$.\\}
    \label{fig:backscatter_isolation}
\end{figure}
The additional noise on the hPD caused by this form of backscattering is mostly white in GEO\,600's shot noise-limited frequency band (see fig.~\ref{fig:backscatter_isolation}).
The coupling is additionally modulated with the sine of the randomly fluctuating mean phase of the stray-light field $\Phi$, this was observed experimentally as a random variation of the noise level on second timescales. This strong modulation of the fast phase-noise coupling spreads out features in the spectrum which therefore appears quite smooth when averaged over many periods of the modulation.\\
\indent The predicted scaling of the backscatter noise with the OPO parametric gain was tested in the detection-band frequency range.
\begin{figure}[ht]
    \centering
    \includegraphics[width=1\textwidth]{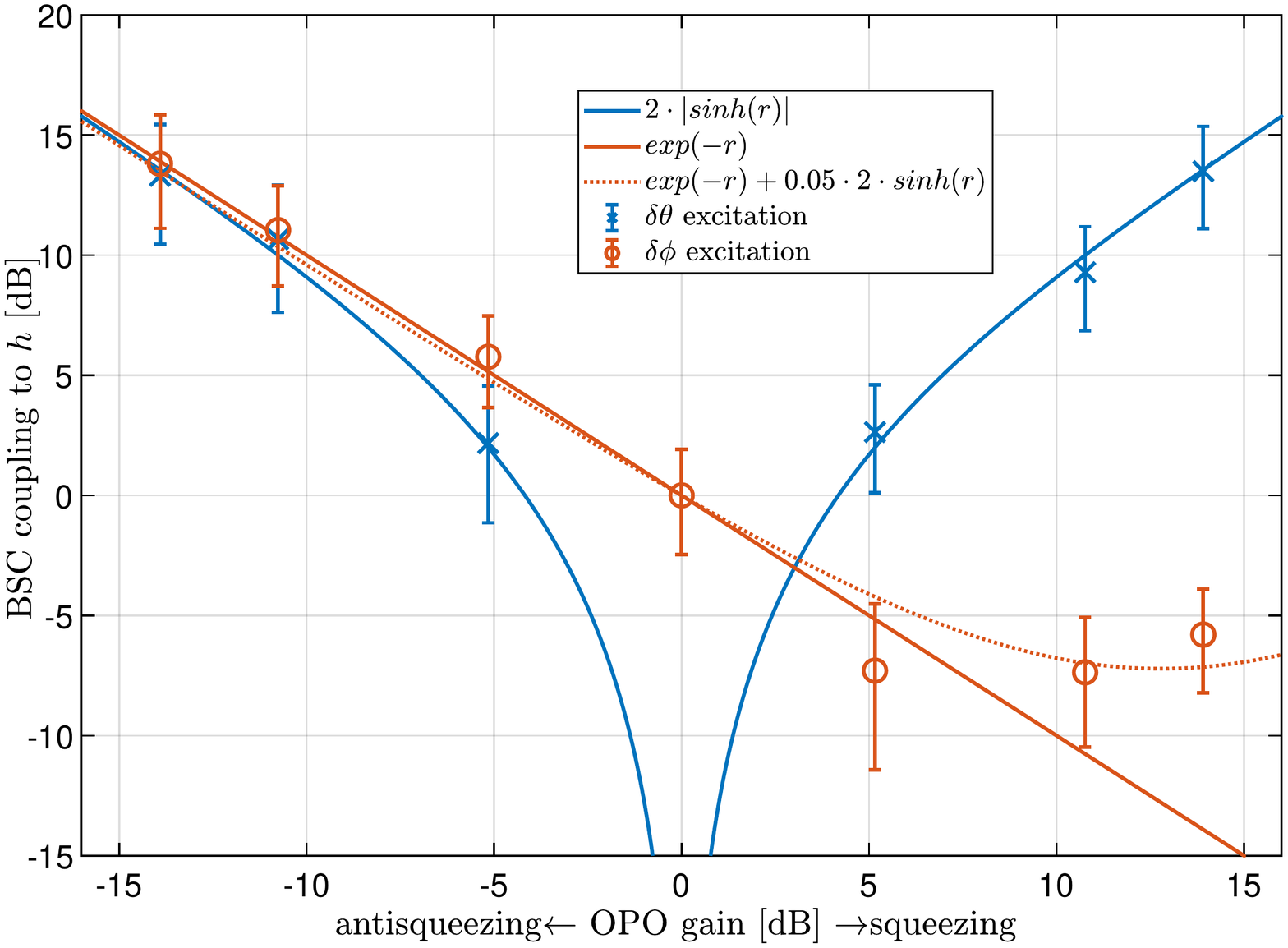}
    \caption
    {Backscatter coupling of fast phase fluctuations as a function of the OPO gain. Negative values mean that anti-squeezing was injected. The data points were obtained by introducing a small phase excitation inside the detection band. The excitation was fed to a piezo mounted mirror and/or to the squeezed light phase actuator via an actuation matrix.
    Each two-minute-long reference time was split into 120 chunks of 1 second each. The backscatter coupling was estimated by determining the BLRMS around the excitation line ($(5500\pm20)\textrm{Hz}$), subtracting the background noise level of the case without excitation. The coupling values were then normalized to the excitation value where no pump light was injected into the OPO. The plot shows the average and the standard deviation of each set of 120 values.}
    \label{fig:coupling}
\end{figure}
To show this dependence experimentally, a harmonic signal with amplitude $\sim30\,\textrm{mrad}$ 
and frequency in the detection band ($5.5\,\textrm{kHz}$) was sent through an actuation matrix to two different actuators. One was a piezo-mounted mirror in the in-air injection path (PZT in fig.~\ref{fig:schematic}), the other was the frequency modulation input of the $80\,\textrm{MHz}$ RF generator used as the local oscillator for the phase locked loop (PLL) of the squeezer master laser with the IFO main laser.
An actuation on the FM port of the RF generator affected only the squeezing angle $\delta\theta$.
 On the other hand, being the injection path traveled by both the squeezed vacuum and the backscattered light, the actuation on the piezo-mounted mirror alone would excite both  $\delta\theta$ and $\delta\phi$ (twice as much).
To excite only the backscattered phase $\delta\phi$, both the actuators were operated at the same time, with amplitudes and phases tuned in order to cancel the modulation on the squeezed vacuum phase. To achieve this, the line at $5.5\,\textrm{kHz}$ in the spectrum of the in-loop error signal for the squeezing phase control was minimized. The latter configuration was similar to what was done in \cite{Wade2013}, but with a different actuation on the squeezing phase: the FM input of the PLL's LO was used instead of a piezo-mounted mirror in the the pump field path.
The different excitations were performed for several values of the pump power and for both squeezing and anti-squeezing applied. Then the resulting band-limited root mean square (BLRMS) of the lines in the readout hPD signal was measured and normalized by the BLRMS of the line for the case where no pump light was injected into the OPO. The result is shown in fig.~\ref{fig:coupling}. 
The predicted models---obtained using the second and third terms of Eq.~(\ref{eq:backscatter_signal_r})---are also plotted, after averaging out the slow backscattered-light phase fluctuations $\Phi$.\\
The excitation induced by the squeezing phase actuation $\delta\theta$ scales as $2|\sinh(r)|$, while the backscattered-light phase $\delta\phi$ excitation should scale as $\exp(-r)$. The dotted line in the figure takes into consideration a non-perfect cancellation of the squeezing angle coupling: $\exp(-r)+0.05\cdot 2\cdot\sinh(r)$.

\section{Backscatter noise reduction by locking the backscattered light phase}\label{sec:BSP_lock}

\begin{figure}[ht]
    \centering
    \includegraphics[width=1\textwidth]{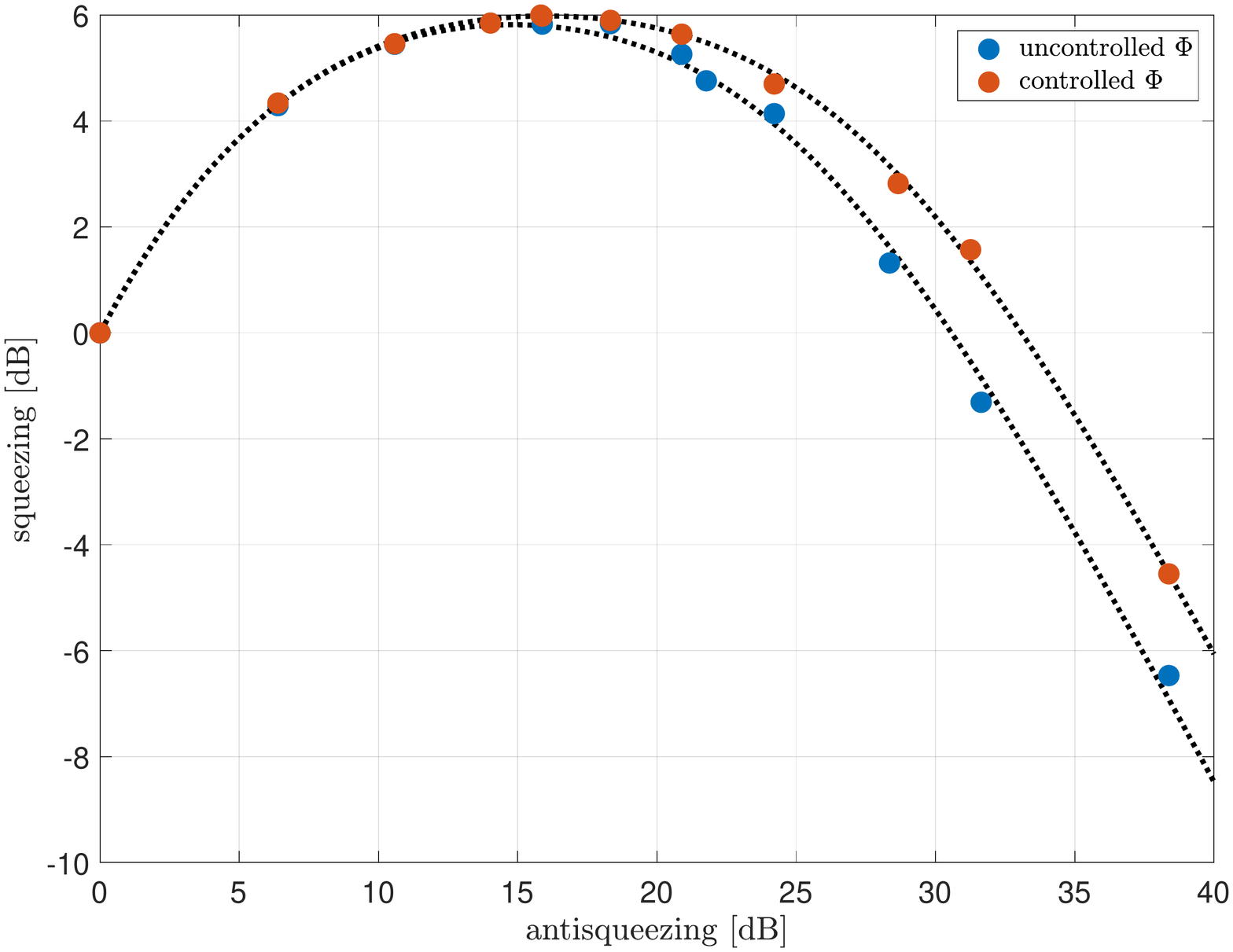}
    \caption
    {Detected squeezing level as a function of measured antisqueezing. The points were obtained by measuring the shot noise level at $5.2\textrm{kHz}$ for both squeezing and antisqueezing injection and dividing it by the shot noise level when no squeezed light was injected. For each antisqueezing level, the squeezing level was measured with both BSP loop on and off. The difference between the two configurations becomes bigger towards higher parametric gain. The dotted traces represent fits of the data points.}
    \label{fig:sqz_vs_asqz}
\end{figure}

To mitigate the effect of backscatter noise there are several general approaches. On the one hand, the amount of stray-light power reaching the squeezer needs to be limited as far as possible.
For this reason, one or more additional Faraday isolators in the injection path (see fig.~\ref{fig:schematic}) are used to suppress this back-propagating light with a typical isolation factor of $40\,\mathrm{dB}$ each \cite{faraday_virgo}. Some small amount of the backscattered light can still reach the OPO, depending mainly on the imperfections of the injection Faraday rotator and on the number and quality of additional Faraday isolators.
On the other hand, due to the linear coupling term in Eq.~(\ref{eq:backscatter_signal_r}), residual fluctuations of the squeezing phase in the detection band must be kept as low as possible: the amount of residual phase noise will depend on the active control and on the intrinsic path stability of the overall optical setup.
Also, the geometry of the OPO plays an important role: if it is built as a ring cavity the stray-light field is intrinsically decoupled from the squeezed vacuum field which propagates in the opposite direction. LIGO OPOs are in a bow-tie configuration for this exact reason \cite{Aasi_2013}. Optical imperfections still lead to coupling of the stray light, but suppressed by over $40\textrm{dB}$ \cite{Chua2011}, which is roughly equivalent to one additional Faraday isolator. 
Finally, the active control of the backscattered light phase presented in this section can be employed in order to minimize the linear coupling of squeezing angle fluctuations.\\
Wade \textit{et al.} \cite{Wade2013} have suggested a technique for backscatter evasion by a combined modulation of the injection phase and pump phase to frequency shift the backscatter noise out of the detection band. The technique was demonstrated in a proof-of-principle experiment, but effective suppression without increasing the RMS phase noise of the squeezing injection appears challenging. A novel active control for the mitigation of backscattered-light-induced noise was developed and employed at GEO\,600: the technique relies on the phase-locking of the backscattered light with the OPO pump field.
The use of this technique makes it possible to achieve a higher level of squeezing, particularly towards a high parametric gain of the OPO. Fig.~\ref{fig:sqz_vs_asqz} shows how the backscattered angle control was an important factor in reaching 6dB of squeezing level at GEO\,600 \cite{Lough2021}. At the optimal antisqueezing level (15.9dB), the squeezing level was improved by 0.2dB (from 5.8dB to 6.0dB) when the backscattering angle control loop was engaged.\\
By looking at Eq.~(\ref{eq:backscatter_signal_r}), a way to mitigate the linear coupling of the residual squeezing angle fluctuations $\delta\theta$ would be to lock the slow backscattered light phase $\Phi$ to $0$ or $\pi$, this corresponds to a parametric deamplification of the backscattered light inside the OPO. The control system was called BSP (Back-Scattering Phase) loop.
The error signal for this control system can be obtained by dithering the pump phase and demodulating the resulting line at the hPD. This signal, which tracks the backscattered-light phase, can be fed back to a piezo-mounted mirror to correct for the injection path length change, thus locking the backscattered light phase to that of the IFO carrier light. The squeezer pump is in turn phase-locked to the IFO carrier field through the coherent control loop \cite{Vahlbruch2006}. Thus, a suitable hierarchy of control bandwidths ensures the phase-locking between backscattered light and pump.\\
The modulation is obtained via PLL. A harmonic signal at frequency above the detection band ($6868\,\mathrm{Hz}$) is fed to the FM port of the $80\,\mathrm{MHz}$ RF generator used as LO for the PLL between the squeezer master laser and the IFO main laser. A trade-off for the amplitude of this modulation must be found, since high values would directly lead to higher RMS phase noise of the squeezed light; on the other hand, the modulation must be strong enough for the line in $h$ to have a reasonable SNR. The optimal modulation amplitude was found to be around $6\,\textrm{mrad}$ at GEO\,600.\\
After rearranging Eq.~(\ref{eq:backscatter_signal_raw}), approximating $\eta_\textup{esc}=1$ and using Eq.~(\ref{eq:pure_state}), the contribution of backscattered light to the photocurrent from the hPD is
\begin{equation}
    \mathrm{I_\textup{bsc}}\propto \sqrt{\mathrm{P_{\textup{IFO}}P_{\textup{bsc,in}}}}\,\Re\left[\cosh r e^{i\phi}-\sinh r e^{i(2\theta-\phi)}\right].
\end{equation}
The squeezing angle $\theta(t)$ is actively kept at $0$ through the coherent control loop, so when a modulation is introduced it becomes $\theta(t)=\gamma\sin(\Omega t)$. For a small modulation depth $\gamma$:

\begin{equation}
    \mathrm{I_{bsc}}\propto\sqrt{\mathrm{P_{IFO}P_{bsc,in}}}\Re\left[\cosh r e^{i\phi}-\sinh r e^{-i\phi}(1+\gamma e^{i\Omega t}-\gamma e^{-i\Omega t})\right].
\end{equation}
When this signal is demodulated with an appropriate demodulation phase $\chi$, the error signal is obtained:

\begin{align}\label{eq:error_signal}
    \epsilon &\propto \sqrt{\mathrm{P_{IFO}P_{bsc,in}}}\,\gamma\,\sinh r\,\cos\chi\sin \phi\\
    &= \sqrt{\delta}\,\mathrm{P_{IFO}}\,\gamma\,\sinh r\,\sin\phi\;,
\end{align}
 which is linear for small values of the backscattered light phase $\phi$. In the second line of the above expression, the linear dependence with factor $\delta$ of backscattered light power and IFO power was considered (at GEO\,600, $\delta=O(10^{-12})$), and the demodulation phase $\chi$ was chosen to maximise the error signal.
 
  \begin{figure}[ht]
    \centering
    \includegraphics[width=1\textwidth]{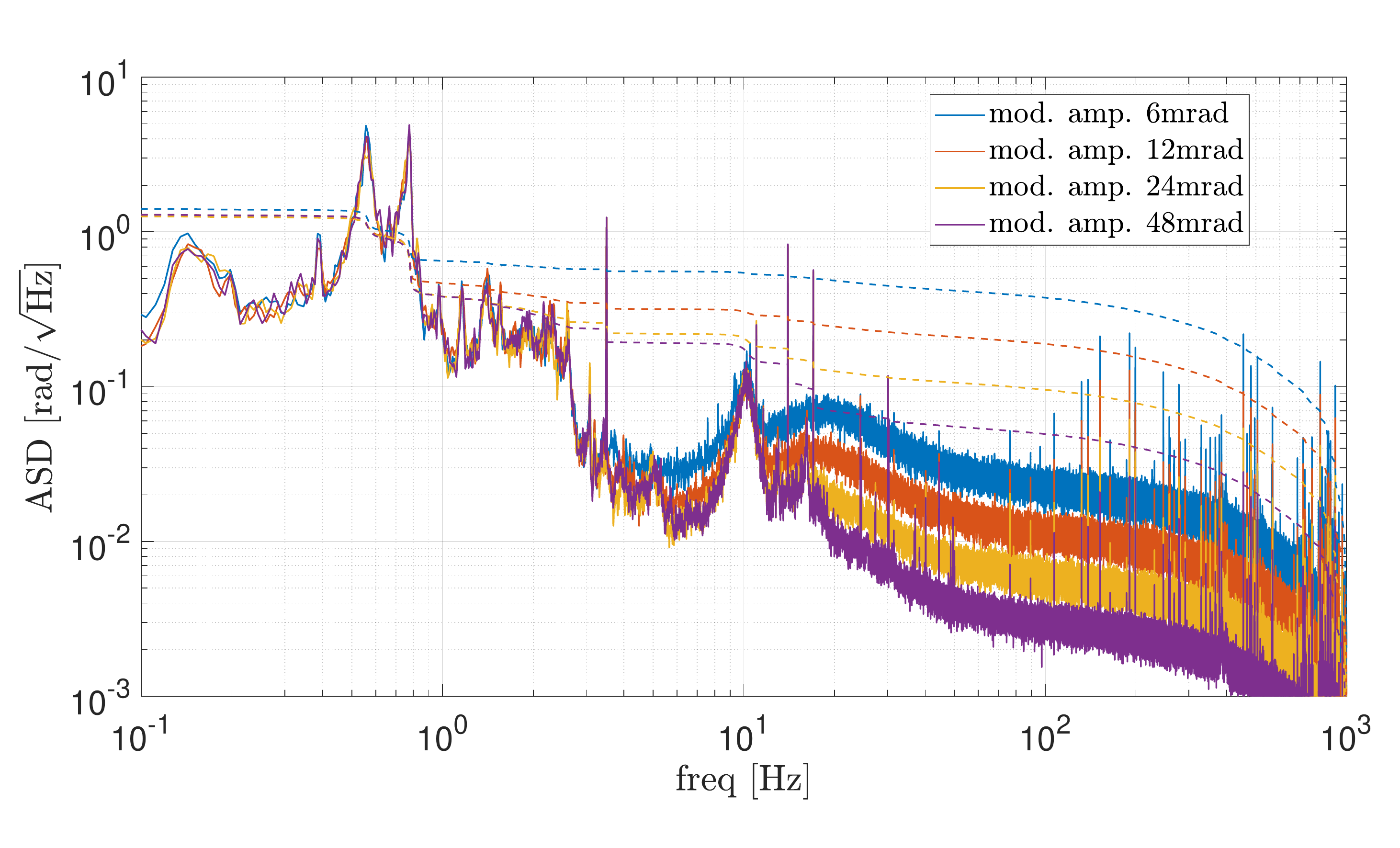}
    \caption
    {Solid traces: spectra of the BSP loop control signal for different amplitudes of the squeezing angle modulation. Dashed traces: cumulative RMS. The resonances below $1Hz$ are due to the suspended optics between the IFO and the OPO. The lines at $3.5\,\mathrm{Hz}$, $11\,\mathrm{Hz}$, $14\,\mathrm{Hz}$, and $17\,\mathrm{Hz}$ are the dithering lines on the BDOs used for the alignment to the OMC \cite{kawabe1994automatic}.}
    \label{fig:asd_feedback}
\end{figure}

\noindent The control signal, obtained after digital filtering of the error signal, is fed to a piezo-mounted mirror (PZT in fig.~\ref{fig:schematic}) in the squeezing injection path. The calibrated amplitude spectral density (ASD) of the control signal (fig.~\ref{fig:asd_feedback}) up to the UGF of the control loop (UGF=$14\,\mathrm{Hz}$) gives an indication of the slow contribution to the backscattered light phase noise, which is due to the length fluctuations of the path between the OPO and the IFO dark port. The major contribution to the RMS of this noise is given by two peaks at $0.55\,\mathrm{Hz}$ and $0.78\,\mathrm{Hz}$, corresponding to the resonance frequencies of suspended optics in the squeezed light path such as the three beam directing optics (BDOs) and the signal recycling mirror (SRM) (see fig.~\ref{fig:schematic}).
The cumulative RMS of the calibrated control signal down to $10\,\mathrm{mHz}$ is $1.3\,\mathrm{rad}$ and it is dominated by the above mentioned suspended optics resonances. In the time domain, the control signal oscillates with an amplitude of $\sim2\,\mathrm{rad}$ indicating that the intrinsic path stability between the OPO and the IFO dark port is good enough to avoid up-conversion of the slow phase fluctuations $\Phi$ into the detection-band frequency range. In fact, the scattering shoulder (described in section \ref{sec:up_convertion}) has a maximum frequency given roughly by the product of the modulation frequency and the modulation amplitude. In the case of the natural oscillation of the squeezing injection path length, obtained from the calibrated BSP control signal, the scattering shoulder would extend up to $\lesssim2\mathrm{Hz}$, below the detection bandwidth of GEO\,600.\\
Above the UGF of the BSP loop, the error signal is dominated by the sensing noise. This corresponds to the shot noise at the hPD photodiode, being the error signal obtained from a demodulation at a frequency belonging to the shot-noise limited band of the detector. Fig.~\ref{fig:asd_feedback} shows how the sensing noise can be reduced by increasing the modulation amplitude ($\gamma$ in Eq.~(\ref{eq:error_signal})) of the squeezing angle. Due to the fact that higher modulation corresponds to higher RMS phase noise of the squeezed light, the loop is operating with a modulation amplitude of $6\mathrm{mrad}$, in order to not degrade the squeezing level. This means that the BSP loop is partially re-injecting the sensing noise in the squeezing injection path longitudinal control. This extra noise can be observed in the coherent control error signal, but it is not a limiting factor for the overall performance of the squeezing angle control.

\section{Influence on phase noise estimation}\label{sec:influence_phase_noise_est}

Relative fluctuations of the angle between the squeezed quadrature and the measured quadrature degrade the shot noise reduction performance by coupling some of the noise from the anti-squeezed quadrature into the measurement quadrature.
Even if the squeezing angle is actively controlled, there will remain some random fluctuations that degrade the squeezing level according to
\begin{equation}
    \frac{\mathrm{RIN_{sqz}}}{\mathrm{RIN_{sn}}}=V_-\cos^2 \theta_\mathrm{RMS} + V_+\sin^2 \theta_\mathrm{RMS}\;,
\end{equation}
where $\theta_\mathrm{RMS}$ is the RMS of the residual phase noise, and
\begin{equation}
    V_\mp = 1\mp\frac{4\eta_{esc}\eta_{inj} x}{(1\pm x)^2 + \left(\frac{\Omega}{\gamma}\right)^2 }
\end{equation}
are the variances of the squeezed and anti-squeezed quadrature \cite[\S2.8]{Dwyer2013a}, relative to the shot noise level. In the last formula, $\gamma$ is the linewidth of the OPO ($2\pi\cdot65\,\textrm{MHz}$ at GEO\,600), and $\Omega$ is the angular frequency at which the measurement is performed. This term can be neglected in the detection-band frequency range. The other parameters were already defined in previous sections.\\
\indent The squeezing phase noise $\delta\theta$ couples to the RIN of the HPD signal also through the mechanism described in section \ref{sec:description_of_bsc}.
If this mechanism is neglected, the phase noise obtained by fitting squeezing level versus anti-squeezing level might be overestimated.
Fig.~\ref{fig:sqz_vs_asqz} shows this kind of plot, the different points were obtained by changing the power of the pump field entering the OPO. The two dotted lines represent the fitting of squeezing versus antisqueezing level obtained when the BSP loop was either engaged or not. The added RIN in terms of shot noise in the case of uncontrolled backscattered-light phase is given by

\begin{equation}
    \mathrm{\frac{RIN_{bsc}}{RIN_{sn}}}=
    \frac{4\eta_{inj}\mathrm{P_{bsc}}}{2\hbar\omega}
     \frac{1}{2}\left[\left( 1-\frac{2\eta_{esc}}{1+x} \right)^2 S_{\delta\phi} + 
    \left(\frac{4\eta_{esc}x}{1-x^2}\right)^2 S_{\delta\theta}
    \right]\,.
\end{equation}
When the BSP loop is engaged, the factor $1/2$ in the above expression has to be replaced with $\sin^2\phi_\mathrm{RMS}\sim\phi_\mathrm{RMS}^2$,
where $\phi_\mathrm{RMS}$ is the RMS of the residual backscattered-light phase noise when $\Phi$ is controlled to be zero.\\
A fit of the data obtained with uncontrolled BSP, performed without considering the backscattered light noise contribution would give the following results:

\begin{align}\label{eq:wrong_phase}
    &\theta_\textup{RMS} = (26.2\pm0.6)\mathrm{mrad}\\
    &\eta_{esc}\cdot\eta_{inj}=(78.3\pm0.8)\%
\end{align}
Where $\eta_{esc}=99\%$.\\
Given $\mathrm{P_{bsc}}$, $S_{\delta\phi}$, and $S_{\delta\theta}$, by fitting the two curves an estimation of $\phi_\mathrm{RMS}$ can obtained, in addition to the injection efficiency $\eta_{inj}$ and the residual squeezing phase noise $\theta_\mathrm{RMS}$.
$\mathrm{P_{bsc}}$ can be obtained from a fit of the backscattering shoulder described by Eq.~(\ref{eq:shoulder}) when no squeeezing is injected ($r=0$).\\
As already pointed out in section \ref{sec:description_of_bsc},  $S_{\delta\phi}$ and $S_{\delta\theta}$ are of similar magnitude at the frequency where the squeezing level is measured, their value can be obtained from the calibrated error point of the squeezing angle control.
The following values were obtained, for $\mathrm{P_{bsc}}=7\,\mathrm{fW}$ and $S_{\delta\phi}=S_{\delta\theta}=10^{-8}\mathrm{rad^2/Hz}$:

\begin{align}
    &\phi_\mathrm{RMS} = (310\pm 20) \mathrm{mrad}\\
    &\theta_\textup{RMS} = (17.8\pm0.8)\mathrm{mrad}\\
    &\eta_{esc}\cdot\eta_{inj}=(78.2\pm0.8)\%\,.
\end{align}
The estimation of the squeezing phase noise thus obtained is in agreement with the value expected from the incoherent sum of known sources: $17\,\mathrm{mrad}$ \cite{Schreiber2018}.
Not considering the effect of backsatter light would lead to an overestimation of the squeezing phase noise, as in Eq.~(\ref{eq:wrong_phase}).

\section{Conclusions}

A theoretical description of the parametric amplification of the backscattered light was given. The model thus obtained provided a description of different coupling mechanisms of the noise introduced by backscattered light when the phase of the squeezed light is actively controlled. The dependence of these coupling mechanisms on the parametric gain was experimentally tested. A scheme to control the phase of the backscattered light by changing the length of the squeezed light injection path was proposed and demonstrated at GEO600. This method has proven to give a significant contribution in maintaining high levels of shot noise reduction, especially at high parametric gain of the OPO and when optical isolation was sub-optimal.\\
\indent The RIN at the IFO readout photodiode caused by the backscatter coupling was described in Eq. (\ref{eq:backscatter_PSD}). This knowledge is a useful tool to inform the design of the squeezing injection path in terms of path stability and optical isolation. 
Of particular interest is the fact that the coupling of squeezing angle fluctuations gets worse for higher parametric gain. This might set even more stringent requirements for the squeezing angle control to achieve the $10\mathrm{dB}$ squeezing level envisioned for future gravitational wave detectors, such as the Einstein Telescope \cite{Hild_2011} and Cosmic Explorer \cite{Reitze2019Cosmic}.

\section{Acknowledgments}

The authors would like to thank Walter Gra{\ss} for his years of expert support in the maintenance of critical infrastructure to the site to include the extensive $400\,\mathrm{m^3}$ vacuum system. The authors are grateful for support from the Science and Technology Facilities Council Grant Ref: ST/L000946/1, the University of Glasgow in the United Kingdom, the Bundesministerium f\"ur Bildung und Forschung, the state of Lower Saxony in Germany, the Max Planck Society, Leibniz Universit\"at Hannover, and Deutsche Forschungsgemeinschaft (DFG, German Research Foundation) under Germany’s Excellence Strategy—EXC 2123 QuantumFrontiers—390837967. This work was partly supported by DFG grant SFB/Transregio 7 Gravitational Wave Astronomy.

\bibliography{sample}

\end{document}